\documentclass[pra,aps,twocolumn, showpacs]{revtex4}

\usepackage{times}
\usepackage{amsmath}
\usepackage{amssymb}
\usepackage{amsthm}
\usepackage{graphicx}
\usepackage{dcolumn}
\usepackage{bm}
\usepackage{multirow}
\usepackage{mathrsfs}


\begin{document}
\title{Distributed quantum information processing with minimal local resources} 

\author{Earl T. Campbell}
\email{earl.campbell@materials.ox.ac.uk}
\affiliation{Department of Materials, Oxford University, Oxford, UK}

\date{Feb, 2006}

\begin{abstract}

We present a protocol for growing graph states, the resource for one-way quantum computing, when the available entanglement mechanism is  highly imperfect.  The distillation protocol is frugal in its use of ancilla qubits, requiring only a single ancilla qubit when the noise is dominated by one Pauli error, and two for a general noise model.  The protocol works with such scarce local resources by never post-selecting on the measurement outcomes of purification rounds.  We find that such a strategy causes fidelity to follow a biased random walk, and that a target fidelity is likely to be reached more rapidly than for a comparable post-selecting protocol.  An analysis is presented of how imperfect local operations limit the attainable fidelity.  For example, a single Pauli error rate of $20\%$ can be distilled down to $\sim 10$ times the imperfection in local operations.

\end{abstract}

\maketitle

\begin{figure}[t]
\centering
\includegraphics{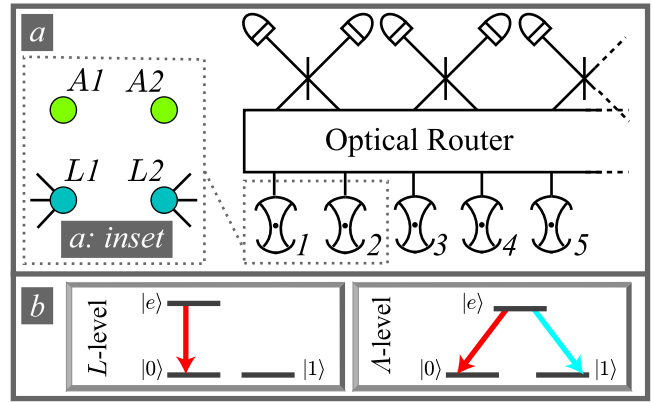}
\caption{An outline of a suitable architecture for a distributed quantum computer.  (\textit{a}) The numbers label local sites, each housing a matter system within a optical cavity.  The optical cavities emit photons into an input port of a multiplexer, which can route any input port to any output port.  Beam-splitters erase which-path information, so that entanglement is generated conditional on the detector signatures.  (\textit{a: inset}) for our primary protocol, each matter system is assumed to have enough level structure to provide two good qubits, a ancilla and a logical qubit. (\textit{b}) two example level structures that would be suitable for the ancilla qubit.  The $L$-level has only one logical state that optically couples to an excited state.  The $\Lambda$-level structure has both levels coupling to a common excited state, however the two transitions are distinguishable by either frequency or polarization.}
\label{fig:Outline}
\end{figure}

The paradigm of {\em distributed} quantum computing (QC) involves a number of simple, optically active structures, each  capable of representing at least one qubit. Relevant examples include trapped atoms \cite{DBMM01a, ODH01a, MMOYMM01a}, and elementary nanostructures such as NV centres within  diamond \cite{BBFM01a, WKN01a, GDPWNJRSGPMTHW01a}. Entanglement {\em between} structures is to be accomplished  through an optical channel, for example by measuring photons after a beam splitter has erased their `which path' information \cite{CCFZ01a, BKPV01a, DBMM01a, DK01a, BK01a, B01a, LBK01a}, as in figure~(\ref{fig:Outline}a).  Remarkably, recent experimental results \cite{MMOYMM01a} demonstrated such an optical channel between ions in separate traps. However, results  to-date show that the `raw' entanglement generated in this way is  liable to have significant noise, well above fault tolerance  thresholds \cite{RHG01a, DHN01a}. Thus it is important to ask, can we exploit the  modest complexity within each local structure in order to {\em  distill} the entanglement to a higher fidelity? 

Originally entanglement distillation was intended for secure quantum communication \cite{BDSW01a, BBPSSW01a, DBCZ01a,  MPPVK01a}, but the same protocols naturally carry over to distributed QC \cite{DB01a, JTSL01a}.  First a noisy entanglement operation produces many noisy Bell pairs between two locations, which these protocols then convert into fewer high-fidelity Bell pairs.  At each local site there must be a certain number of qubits available, one logical qubit that is directly involved in the computation, and some number of ancilla qubits.  Computation is performed by distilling a high-fidelity Bell pair between two ancilla qubits, and then using it to implement a gate between two logical qubits.  In addition to allowing purification, ancilla qubits protect the logical qubits against damage from probabilistic gates \cite{DB01a, DBMM01a, BBFM01a, JTSL01a}.  Since these protocols emphasize implementing a good fidelity gate, we refer to them as \textit{gate-based} protocols.  For significant purification of noise from a depolarizing source, these proposals require $3$ or $4$ ancillary qubits \cite{DB01a, JTSL01a}; whereas for phase noise, the number of ancillas can be reduced by one \cite{JTSL01a}.  

Another family of distillation protocols emerged after the one-way model of quantum computing showed that all the entanglement necessary for computation is present in a class of states called graph, or cluster, states \cite{RBB01a, RB01a, HDERNB01a, HEB01a}.  The distillation of graph states is akin to error correction, as it consists in repeated measurement of the stabilizers that describe the graph state.  In virtue of this feature, we refer to these as \textit{stabilizer-based} protocols. The first such protocol uses noisy copies of a graph state and post-selects upon detection of a single error \cite{DAB01a, ADB01a, KMBD01a}.  Further proposals cast aside the need for post-selection at the cost of a stricter error threshold \cite{GMR01a}, above which, distillation is possible.  These proposals use a combination of noisy copies of the graph state and highly purified GHZ states.  Because of the size of the entangled states in the ancilla space, iterating these distillation protocols may take longer than for gate-based protocols.  A significant temporal overhead will occur when the entangling operation has a high failure rate.  Building large entangled states in the ancilla space also restricts the class of employable entangling operations, excluding entangling protocols that only produce Bell pairs \cite{DBMM01a, DK01a}.  Of course, provided we have enough local qubits to provide ancillas for our ancillas, these disadvantages are easily nullified.  However, many systems which are potentially well suited for distributed QC may be very limited in the number of qubits they can embody.

By blending ideas from the gate-based and stabilizer protocols, this paper proposes an entanglement distillation protocol which performs rapidly whilst requiring fewer ancillas than previous protocols.  The bulk of this paper shows that one ancilla is sufficient to distil errors from dephasing noise.  We then extend the protocol to cover depolarizing noise; as with other schemes this requires an additional ancilla, which we use to reduce the depolarizing noise to a dephasing noise.  Like gate-based protocols, we build up a graph state edge-by-edge, with ancillas never building entangled states larger than Bell pairs.  However, as with stabilizer-based protocols, our proposal repeatedly makes stabilizer measurements directly onto the qubits constituting the graph state.  Ancillas must typically be optically active, such as in an $L$ or $\Lambda$ level configuration (see Fig.~\ref{fig:Outline}b).  Qubits are laBelled $Ax$ and $Lx$ for ancilla and logical qubit, respectively, at local site $x$.

First our analysis will focus on the case when the noisy entanglement channel is dominated by one type of Pauli error, which may be very  severe.  Without loss of generality, we describe the channel as being affected by phase noise, such that two ancillas $A1$ and $A2$,  can be put in the mixed state: 
\begin{equation}
\label{eqn:channel}
	\rho_{A1, A2} = ( 1 - \varepsilon ) \vert \Psi_{+} \rangle \langle \Psi_{+} \vert + \varepsilon Z_{A} \vert \Psi_{+} \rangle \langle \Psi_{+} \vert Z_{A} , 
\end{equation}
where $Z_{A}$ is the Pauli phase-flip operator acting on either $A1$ or $A2$, and $\vert \Psi_{+} \rangle = \vert 0 \rangle_{A1} \vert 1 \rangle_{A2} + \vert 1 \rangle_{A1} \vert 0 \rangle_{A2}$.  If the dominant noise is a different Pauli error, or different Bell pairs are produced, then local rotations can always bring the state into the form of equation ~\ref{eqn:channel}.  Furthermore, only a single $Z$ error is possible as this Bell state is invariant under the bilateral $Z_{A1}Z_{A2}$ rotation.  Scenarios where such a noise model may arise include parity based entangling operations \cite{BK01a, DK01a} that possess a degree of robustness against bit-flip errors.  

\begin{figure}[t]
\centering
\includegraphics{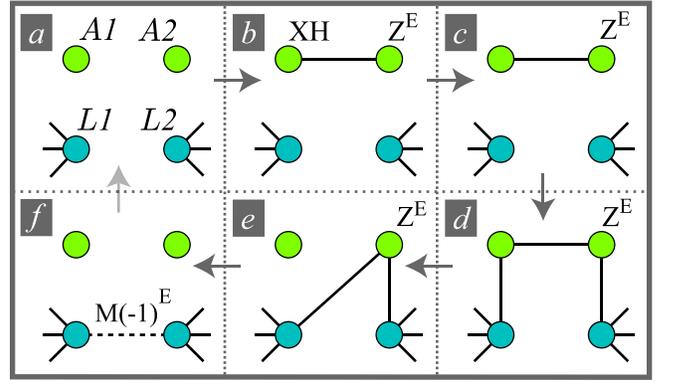}
\caption{The sequence of operations required to pump entanglement down to two logical qubits $L1$ and $L2$, shown in the graph state notation.  (\textit{a}) Ancillas are prepared in the $(\vert 0 \rangle + \vert 1 \rangle)/\sqrt{2}$ state, and the logical qubits are part of some larger graph state $\vert \mathscr{G} \rangle$; (\textit{b}) An entangling operation is performed between the ancillas, with the possibility of $Z$ noise; (\textit{c}) A $H \cdot X$ is applied to ancilla $A1$; (\textit{d}) At both local sites control-$Z$ operations are performed between ancilla and logical qubit; (\textit{e}) Ancilla $A1$ is measured in the $Y$ basis; (\textit{f}) Ancilla $A2$ is measured in the $X$-basis.  The possibility of a $Z$ error is tracked by using $Z^{E}$, where $E=1$ tracks an error, and $E=0$ tracks the errorless state.  The measurement outcome is represented by $M$, where $M=+1$ for a $\vert 0 \rangle$ measurement and $M=-1$ otherwise.  The dotted line between the logical qubits represents a projection operator $P_{\pm}$ between the logical qubits, where the sign is equal to $M(-1)^{E}$.}
\label{fig:NoisyParity}
\end{figure}

After producing noisy entanglement between two ancillas, the entanglement is pumped down to the logical qubits, resulting in a quantum operation on the logical qubits.  The target (perfect) entangling operation we aim to eventually achieve is either of the parity projections:
\begin{eqnarray}
	P_{-} & = &2 ( \vert 01 \rangle \langle 01 \vert + \vert 10 \rangle \langle 10 \vert ), \\ \nonumber
	P_{+} & = &2 ( \vert 00 \rangle \langle 00 \vert + \vert 11 \rangle \langle 11 \vert ),
\end{eqnarray}
which act on the logical qubits $L1$ and $L2$, and have an additional normalization factor of 2 that simplifies later expressions.  The only assumption we make about the initial state of the logical qubits is that they are part of a graph state (in the constructive definition), such that they have equal magnitude in both parity subspaces, $\langle \mathscr{G} \vert P_{+} \vert \mathscr{G} \rangle = \langle \mathscr{G} \vert P_{-} \vert \mathscr{G} \rangle$; where $\vert \mathscr{G} \rangle$ denotes the graph state of all the logical qubits.  Both $P_{-}$ and $P_{+}$ allow arbitary graph growth, and which projection we eventually obtain is unimportant as $P_{\pm} \vert \mathscr{G} \rangle$ differ only by local rotations \footnote{$P_{+} \vert \mathscr{G} \rangle = X_{L1} ( \bigotimes_{j \in N(L1)} Z_{j}) P_{-} \vert \mathscr{G} \rangle $, where $N(L1)$ is the set of neighbours of $L1$.  }.  We will see that entanglement distillation results from repetition of an entanglement transfer procedure.

Each round of our protocol is an entanglement pumping procedure, described graphically in Fig.~(\ref{fig:NoisyParity}). Every round of purification begins with performing a noisy entangling operation between two ancillas, $A1$ and $A2$.  This operation may be probabilistic provided that success is \textit{heralded},  in which case it is repeated until successful.  Next, a series of local operations must be performed.  First, apply a bit-flip then Hadamard to one ancilla, say $A1$, and then two control--Z operations between each ancilla and its logical qubit.  The resulting state is a graph state, with the possibility of a $Z$ error on $A2$.  Then, measure $A1$ in the $Y$-basis (correcting for any by-product), giving the state described by Fig.~\ref{fig:NoisyParity}e.  Finally, $A2$ is measured in the $X$-basis.  When no error was present, this measurement performs a parity check on the logical qubits, $L1$ and $L2$; that is, we measure the observable $Z_{L1}Z_{L2}$.  On the first round of pumping, the odd and even parity outcomes will occur with 50/50 probability.  Accounting for the possibility of a $Z$ error causes a noisy parity measurement, or quantum operation:
\begin{equation}
\label{eqn:Qop}
	\mathcal{P}_{\Delta} \left(  \vert \mathscr{G} \rangle \langle \mathscr{G} \vert  \right) = \frac{\alpha^{\Delta} P_{+} \vert \mathscr{G} \rangle \langle \mathscr{G} \vert P_{+} + \alpha^{-\Delta} P_{-} \vert \mathscr{G} \rangle \langle \mathscr{G} \vert P_{-} } {\alpha^{\Delta}+\alpha^{-\Delta}},
\end{equation}
where $\alpha=(\varepsilon^{-1}-1)^{\frac{1}{2}}$, and $\Delta=M1$ is $+1$ for a $\vert 0 \rangle$ measurement outcome, and $-1$ for $\vert 1 \rangle$.

If we repeat the entanglement pumping procedure $n$ times, then we will get a series of measurements results $M1$,$M2$,$\ldots Mn$.  Concatenating the mapping, for each measurement result, we get back an operation of the same form but with $\Delta= \sum_{i=1}^{i=n} Mi$.  The core of our proposal is that we continue to purify the qubits until $|\Delta|$ reaches some value $\Delta_{H}$ at which point we halt the procedure.  $\Delta_{H}$ is chosen such that it corresponds to a target fidelity $F_{T}$, where the fidelity is simply $F(\Delta)=(1+ \alpha^{-2| \Delta |})^{-1}$.  

\begin{figure}[t]
\centering
\includegraphics{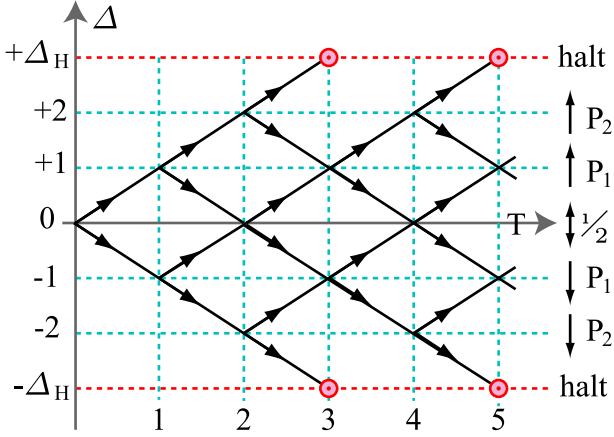}
\caption{The evolution of $\Delta$ against $T$, the number of rounds of entanglement pumping.  The state evolves as a biased random walk in $\Delta$.  The weighting of probabilities at a some point $\Delta$ is such that there is a probability $P_{|\Delta|}$ for increasing the magnitude of $\Delta$, where $P_{|\Delta|}$ is defined in equation \ref{eqn:PD}.  The red dashed lines represent the halting lines for $\Delta$, and in this example $\Delta_{H}=3$.  Note that, the paths to the halting line, occur at $T=\Delta_{H}+2n$, for non-negative integer $n$.}
\label{fig:RandomWalk}
\end{figure}

In contrast to previous gate-based protocols, our protocol is not post-selective (NPS).  Analogous post-selecting protocols (PS) using an equivalent entangling pumping procedure to eliminate phase errors have already been proposed \cite{JTSL01a}.  The essential difference for PS is that it resets upon any measurement outcome, $Mx$, that differs from the first measurement outcome, $M1$; a reset consists in measuring out the qubits being distilled, bringing them back to $\Delta=0$.  Benefits of NPS are two-fold: (\textit{i}) since purification is never restarted it is safe to operate directly on the logical qubits, hence we eliminate the need for an additional ancilla that exists in PS protocols; (\textit{ii}) the probability of success within $T$ rounds is never less than for PS, indeed, we shall show that NPS significantly outperforms PS in this regard.  A point in favour of PS is that, if can freely use multiple ancillas, then PS may achieve a higher asymptotic limit of fidelity (due to the effect of errors in local operations).  However, we shall see that NPS can still attain fidelities within fault tolerance thresholds.

\begin{figure}[t]
\centering
\includegraphics{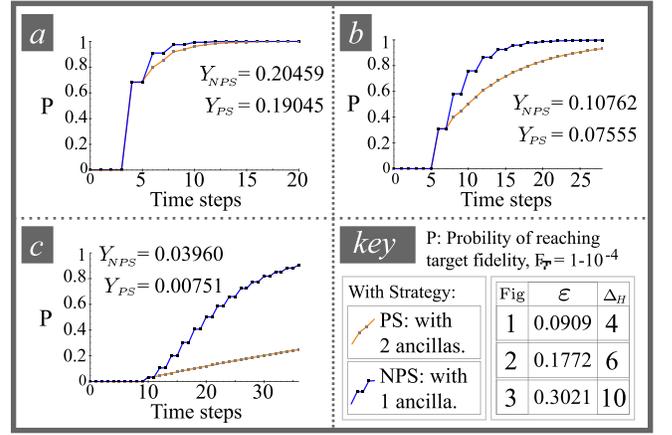}
\caption{A comparison of the rapidity of our proposal versus a post-selection protocol, with a target fidelity $F_{T} = 1 - 10^{-4}$.  The three plots represent different values for the error rate $\varepsilon$, and hence require a different value of $\Delta_{H}$; shown in key.  On each plot we show the probability of success against the number of rounds $T$, for both the protocol NPS (blue) and PS (orange).  The yield for NPS and PS ($Y_{nps}$ and $Y_{ps}$) is given on each plot.  Notice that for NPS, the probability of success increases in steps.  This is explained by figure \ref{fig:RandomWalk}, which shows that successful paths are separated by 2 time steps.}
\label{fig:ConstFidelity}
\end{figure}

Returning to our consideration of the evolution of $\Delta$ in our protocol, it is clear that at each purification step, $T$, $\Delta$ can either increase or decrease by 1.  Hence, the evolution bears similarities to a random walk, illustrated by Fig.\ref{fig:RandomWalk}.  It differs from a random walk in two regards: (\textit{i}) it halts when it reaches $\Delta= \pm |\Delta_{H}|$; (\textit{ii}) the probabilities are the biased when $\Delta \neq 0$.  The bias increases the chance of walking in the direction of larger $|\Delta|$, which occurs with probability:
\begin{equation}
\label{eqn:PD}
	P_{D} = \frac{(1- \varepsilon) \alpha^{D} + \varepsilon \alpha^{-D}  }{\alpha^{D}+ \alpha^{-D}},
\end{equation}
where $D=|\Delta|$. On the face of it, it seems that the probability of walking to a state $\Delta$ in $T$ steps is dependent upon which path is taken.  However, the probability of a kink in the path --- $D$ increasing and subsequently decreasing --- is independent of $D$, and is $k = P_{D} (1- P_{D+1}) = \varepsilon (1- \varepsilon)$. Hence, each path occurs with probability:
\begin{equation}
	P \text{path} (D, T) = \left( \prod^{D-1}_{d=0} P_{d} \right) k^{\left( \frac{T-D}{2} \right)}.
\end{equation}
The total probability of walking to ($D, T$) is the product of $P\rm{path}$($D, T$) with the number of paths to that position.

We have calculated the total probability of success, after $T$ rounds, by summing over all the different ways of reaching the halting line.  For comparison, we performed the analogous calculation for an otherwise equivalent PS protocol.  Figure \ref{fig:ConstFidelity} shows PS and NPS protocols for a target fidelity of $F_{T}=1-10^{-4}$, with each plot being for a different error rate $\varepsilon$. Note that, for higher error rate or higher target fidelity, the width of the random walk is wider (larger $\Delta_{H}$).  In this regime, the superiority of NPS increases, as more entanglement can be lost upon post-selection.  Conversely, when $\Delta_{H}=2$ the protocols are effectively identical, as stepping back will take the walk to the origin.  A protocol's \textit{yield} is the expected ratio of distilled Bell pairs to used noisy Bell pairs.  Since each time step uses a Bell pair, the yield of a protocol is $1/\langle T \rangle$, with some values given in Fig.~\ref{fig:ConstFidelity}.  Since NPS is a faster protocol than PS, it also has a superior yield.

\begin{figure}[t]
\centering
\includegraphics{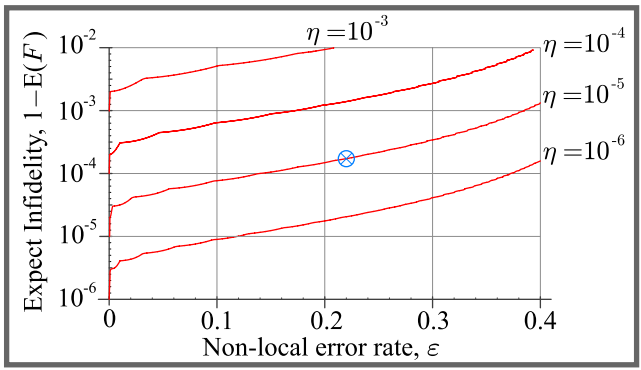}
\caption{A logarithmic plot of the expected infidelity, $1-E(F)$, when additional error sources affect our distillation protocol.  The plot is a function of $\varepsilon$, the probability of a phase error occurring in the long-range entangling operation.  Each curve is a different value of $\eta$, the probability per distillation round that some other error occurs.  The blue cross marks an example discussed in the text, where the dephased Bell pairs are obtained by prior distillation on raw depolarized Bell pairs.}
\label{fig:ExpectedFidelity}
\end{figure}

For this idealized error model we can asymptotically approach unit fidelity.  However, it is important to consider how other errors limit the maximum attainable fidelity.  For simplicity, we take an aggressive error model where if a single error occurs once, then the overall entangling gate has fidelity zero.  We use $\eta$ to denote the probability per time step that an error occurs, where these errors can result from either faulty local operations or noise in the entanglement channel that is orthogonal to the distilled dominant noise.  Furthermore, we approximate the chance of an error after $T$ rounds by its upper bound, $\eta T$.  Once we reach $\Delta_{H}$ the fidelity will depend  on the number of time steps taken.  Therefore, we calculate the expectation of the fidelity, $E(F)$.  Figure~\ref{fig:ExpectedFidelity} shows how the expected infidelity, $1 - E(F)$, varies with $\varepsilon$ and $\eta$.  Since the optimal choice of $\Delta_{H}$ changes with $\varepsilon$ this produces inverted humps along the curves, which are more pronounced for small $\varepsilon$.  On all curves the behaviour is roughly  the same; we can characterize the performance by noting that when the  dominant error rate is 0.2 (i.e. 20\%), and the probability of other  error sources is $\eta$, then the protocol brings {\em all} error probabilities to order $10\eta$.  Given that relevant fault tolerance strategies can handle noise of order $1\%$ \cite{RHG01a, DHN01a}, the single-ancilla distillation may suffice when $\eta$ is of order $0.1\%$. 

If the orthogonal errors are too large, then an additional ancilla
is required. Using the two ancillas, bit-errors are first
distilled away by a post-selective protocol, such as in \cite{JTSL01a}.
Orthogonal errors are at their most extreme when the noise
is depolarizing, producing Werner states of fidelity $F_{0}$. As
an example, we consider the distillation of Werner states of
$F_{0} = 0.85$; a rigorous analysis is provided in appendix \ref{sec:App}. Five
rounds of distillation reduces the orthogonal errors to $\sim 10^{-5}$,
after which the phase noise has accumulated to $\varepsilon = 0.22$; we
have not gained fidelity, but we have mapped all noise into
phase noise. These dephased Bell pairs are used in our primary
distillation protocol, and are distilled to $E(F) \sim 10^{-4}$.
This result is marked on Fig.~\ref{fig:ExpectedFidelity} , where $\eta$ is taken to equal the
remaining non-phase errors. As a final remark, note that if we
create GHZ states among ancillas (rather than Bell-pairs) then
our strategy can be combined with the band-aid protocol of
\cite{GMR01a}, increasing tolerance of imperfect local operations.

In conclusion, using fewer ancillas than previous proposals, an otherwise intolerably-large error is rapidly reduced below error-correction thresholds \cite{RHG01a, DHN01a}.  For dephasing/depolarizing noise models, the protocol needs only one/two ancillas, respectively.  The author thanks Simon Benjamin, Joseph Fitzsimons, Pieter Kok, and Dan Browne for useful discussions.  This research is part of the QIP IRC (GR/S82176/01).

\appendix

\section{Converting Werner states into Dephased Bell states.}
\label{sec:App}

In the main article we propose a protocol that uses a single extra ancilla to distil a graph state in an \textit{edge-by-edge} fashion, provided an entanglement channel that generates dephased Bell states.  However, if the entanglement channel suffers depolarizing noise then this generates a rank-4 Bell diagonal mixed state, known as a Werner state. Here we show that with an additional ancilla a well-known post-selective protocol can be used to convert these Werner states into dephased Bell states.  As an example, we calculate the accumulated phase noise when the entanglement channel produces Werner states with a fidelity of $F_{0}=0.85$, and non-phase errors must be reduced to order $10^{-5}$.

Given an entanglement channel that generates Werner states of the form:
\begin{eqnarray}
	\rho & = &  F_{0} \vert \Phi^{+} \rangle \langle \Phi^{+} \vert  +  \left( \frac{1-F_{0}}{3} \right) \bigg[ Z \vert \Phi^{+} \rangle \langle \Phi^{+} \vert Z \\ \nonumber
	     &  & + X \vert \Phi^{+} \rangle \langle \Phi^{+} \vert X + Y \vert \Phi^{+} \rangle \langle \Phi^{+} \vert Y \bigg] ,
\end{eqnarray}
where $\vert \Phi^{+} \rangle = \vert 00 \rangle + \vert 11 \rangle$.  The $X$ and $Y$ noise can be distilled by performing repeated noisy measurements of the ZZ observable.  This is implemented by first performing bilateral control-phase rotations, with both controls from one EPR pair, and both targets on another EPR pair.  The control qubits are measured in the X-basis, and we post-select on the even parity measurement outcome.  After $n$ successful rounds the target qubits are in some Bell diagonal state:
\begin{eqnarray}
	\rho_{n} & = & a_{n} \vert \Phi^{+} \rangle \langle \Phi^{+} \vert +  b_{n} X \vert \Phi^{+} \rangle \langle \Phi^{+} \vert X \\ \nonumber 
	         &   & + c_{n} Z \vert \Phi^{+} \rangle \langle \Phi^{+} \vert X + d_{n} Y \vert \Phi^{+} \rangle \langle \Phi^{+} \vert Y 
\end{eqnarray}
After another successful round of distillation, the state is transformed such that:
\begin{eqnarray}
	\rho_{n+1} & \propto & F_{0} P_{+} \rho_{n} P_{+} + \left( \frac{1-F_{0}}{3} \right) \bigg[  P_{-} \rho_{n} P_{-} \\ \nonumber & & + Z P_{+} \rho_{n} P_{+} Z + Z P_{-} \rho_{n} P_{-} Z   \bigg]
\end{eqnarray}
Where each term comes from considering a different Bell state contribution to the mixed state of the control qubits.    Contributions with $Z$ or $Y$ noise generate projections into the opposite parity space, as these errors anti-commute with one of the X-basis measurements.  Contributions with $X$ or $Y$ noise will result in a phase error on a target qubit.  These errors propagate down to the target qubits because $X$ and $Y$ rotations do not commute with control phase gates, rather these rotations change the gate's control from the $\vert 1 \rangle$ state to $\vert 0 \rangle$.  Before normalization, the density matrix coefficients obey the recursive relations:
\begin{eqnarray}
	a_{n+1} & = &  F_{0} a_{n} + \left( \frac{1-F_{0}}{3} \right)  c_{n}\\
	c_{n+1} & = &  \left( \frac{1-F}{3} \right) a_{n} + F c_{n} , \\
	b_{n+1} = d_{n+1} & = & \left( \frac{1-F_{0}}{3} \right) (b_{n}+ d_{n})
\end{eqnarray}
Fixing the $n=1$ coefficients to those of our undistilled Werner state, we can derive:
\begin{eqnarray}
	a_{n} + c_{n} = \left(  \frac{1+2F_{0}}{3} \right)^{n} \\
	b_{n} = d_{n} = \frac{1}{2} \left( \frac{2}{3} (1-F_{0}) \right)^{n}
\end{eqnarray}
Hence, after $n$ rounds of post-selective distillation, the remaining $X$ and $Y$ noise has a magnitude:
\begin{eqnarray}
	N_{XY}(F_{0},n) & = & \frac{b_{n}+d_{n}}{a_{n}+b_{n}+c_{n}+d_{n}} \\ \nonumber
	 & = & \left( 1 + \left( \frac{1 + 2F_{0}}{2-2F_{0}} \right)^{n} \right)^{-1} 
\end{eqnarray}
To give a numerical example, consider a source of Werner states of fidelity $0.85$, and we wish to reduce $N_{XY}$ to order $10^{-5}$. It is easy to calculate that five rounds of distillation is sufficient since $N_{XY}(0.85, 5)= 1.69 \cdot 10^{-5}$.  

Finally, we need to calculate how the undistilled phase noise has changed through these five rounds of distillation.  From the recursive relations, an unnormalized form of $c_{n}$ can be derived:  
\begin{equation}
	c_{n} = \frac{1}{2} \left[ \left( \frac{1+2F}{3}  \right)^{n} - \left( \frac{4F-1}{3}  \right)^{n}  \right]
\end{equation}
After normalization, we have the remaining phase noise:
\begin{eqnarray}
 N_{Z}(F_{0}, n) & = & \frac{c_{n}}{a_{n}+b_{n}+c_{n}+d_{n}} , \\
 N_{Z}(0.85,5)& = & 0.22
\end{eqnarray}
It is an interesting feature of distillation that although this state has a lower fidelity than the Werner state, it is a more useful resource for the subsequent level of distillation.  Hence, fidelity alone is a poor indicator of the distillable entanglement of a mixed state.

\end{document}